

\input amstex 

\magnification 1200 \tolerance 5000\documentstyle{amsppt}

\define\hbr{\hbar}

\define\A{\Cal A}

\define\C{\Bbb C}
\define\cb{\bold c}

\define\Gg{\Bbb G}
\define\Gc{\Cal G}
\define\g{\goth g}
\define\Hc{\Cal H}

\define\Hg{\goth H}

\define\k{\Bbbk}

\define\R{\Bbb R}

\define\undn{\underline n}
\define\undx{\underline x}

\define\Z{\Bbb Z}

\define\De{\Delta}

\define\la{\lambda}

\define\varep{\varepsilon}

\define\id{\operatorname{id}}
\define\Imm{\operatorname{Im}}

\define\re{\operatorname{Re}}

 \define\vrt#1{\vert#1\vert}
 \define\pd#1#2{\dfrac{\partial#2}{\partial #1}}
 \define\Utr{{}^{t}U^{-1}} \define\cc#1{c_{(#1)}}

 \topmatter \title Quantum Heisenberg groups and Sklyanin algebras
 \endtitle \author Nicol\'as Andruskiewitsch, Jorge Devoto and
 AlejandroTiraboschi
\endauthor

\address N. Andruskiewitsch: Max-Planck-Institut f\"ur Mathematik.
Gottfried-Claren Strasse 26.  5300 Bonn 3.  Germany\newline Permanent
address: FAMAF. Valpara\'\i so y R. Mart\'\i nez. 5000 Ciudad
Universitaria. C\'ordoba.  Argentina \endaddress \email andrus\@
mpim-bonn.mpg.de\endemail

\address J. Devoto: International Centre of Theoretical Physics. P.O.
Box 582. (34100) Trieste -- TS.  Italy\endaddress \email devoto\@
ictp.trieste.it \endemail

\address A. Tiraboschi: International Centre of Theoretical Physics.
P.O. Box 582. (34100) Trieste -- TS. Italy\newline Permanent address:
FAMAF. Valpara\'\i so y R. Mart\'\i nez.  5000 Ciudad Universitaria.
C\'ordoba. Argentina \endaddress \email tirabo\@ ictp.trieste.it
\endemail

\abstract We define new quantizations of the Heisenberg group by
introducing new quantizations in the universal enveloping algebra of
its Lie algebra. Matrix coefficients of the Stone--von Neumann
representation are preserved by these new multiplications on the
algebra of functions on the Heisenberg group. Some of the new
quantizations provide also a new multiplication in the algebra of
theta functions; we obtain in this way Sklyanin algebras.
 \endabstract

 \thanks N.A.: Forschungsstipendiat der Alexander von
 Humboldt-Stiftung.\quad J.D. and A.T.: Postdoctoral fellowship,
 ICTP.\quad N.A. and A.T: This work was also partially supported by
 CONICET, CONICOR and FAMAF (Argentina). \endthanks \endtopmatter

\document

\subheading{\S 1. Introduction}It is known that theta functions arise
as certain matrix coefficients of the Stone--von Neumann
representation (see for example \cite{Mu}). We are interested in the
algebra of thetas, i.e.  the homogeneous coordinate ring \cite{Mu,
  Section 10}.  It is easy to see that the multiplication of thetas
corresponds to the multiplication of matrix coefficients in the
algebra of functions on the Heisenberg group.  The purpose of this
paper is to state a quantum analogue of this phenomenon.  We introduce
new quantizations of the Heisenberg group; some of them give rise to
quantum mutiplications on the ring of theta functions. We obtain in
this way Sklyanin algebras. (See \cite{Sk}, \cite{ATV}, \cite{SS},
\cite{LS}, \cite{OF}).

\medpagebreak
Besides the so-called quantum Heisenberg algebra (known to
physicists for a long time, see \cite{Ku}, \cite{JBS},
\cite{GF}, \cite{R}), a quantized enveloping algebra
$U_\hbr(\g)$ of the Heisenberg-Lie algebra was introduced in
\cite{Ce et al}, and (independently but later) in
\cite{ALT}.  We discuss $U_\hbr(\g)$ in Section 2. The
novelty is that we determine the primitive spectrum of $\Bbb
C_\hbr[G]$ (the algebra of functions on the quantum
Heisenberg group) and constate a bijective correspondance
with the set of symplectic leaves of the corresponding
Poisson structure (previous work in the semisimple case was
done in \cite{VS},\cite{LS},\cite{HL},\cite{J}). It is
however more convenient for our purposes to work with other
quantizations of the Heisenberg group.

\medpagebreak We recall basic facts about  the Stone--von
Neumann theorem in Section 3. Let us fix a positive integer
$m$. Then any irreducible unitary representation of the
Heisenberg group, such that its center (a copy of $S^{1}$)
acts by $z\mapsto z^{m}\id$, is the tensor product of a trivial
representation with $\Hc^{(m)}$, the Stone-von Neumann
representation of weight $m$. Applying this to the tensor
product $\Hc^{(m)}\otimes \Hc^{(p)}$, we see that the
product of a matrix coefficient of $\Hc^{(m)}$ and a matrix
coefficient of $\Hc^{(p)}$ is a matrix coefficient of
$\Hc^{(m + p)}$. Here the product of matrix coefficients
can be thought in the dual of $U(\g)$, the enveloping
algebra of the Heisenberg--Lie algebra. Clearly, different theta
functions arise as matrix coefficients
of the various $\Hc^{(m)}$.

So our first
approximation is to introduce a new comultiplication in
$U(\g)[[\hbr]]$ with the same property. We do this in
section 4. We also explain carefully there how more new
comultiplications in $U(\g)[[\hbr]]$, preserving
coefficients of the Stone--von Neumann representations,
should look like. In section 6 we present these
 new comultiplications. They provide quantizations of the
Heisenberg--Lie group in the sense of \cite{B et al},
\cite{Dr1}.   In Section 5, we discuss our new
multiplication (on the algebra of functions) in a purely
algebraic setting, which encompass also the presentation of
Skylanin algebras given in \cite{ATV}. We show in Section 7
that some of the comultiplications introduced in Section 6
give rise to ``quantum" multiplication in the ring of theta
functions, via the identification of the later with
particular matrix coefficients. Thanks to the results of
Section 5, we see that we have obtained Sklyanin algebras.
On another direction, we show in Section 8 that the quantum
Heisenberg algebra mentioned above is a braided Hopf
algebra.

\subheading{\S 2} We want first to discuss the algebra of functions on
the quantum Heisenberg group considered in \cite{Ce et al},
\cite{ALT}. Let $\hat \g$ (resp., $\g$) be the extended Heisenberg Lie
algebra, i.e.  the Lie algebra spanned by $x_{i}$, $y_{i}$, $1\le i
\le g$, $z$, $d$ with brackets $$\gather [x_{i},y_{j}] = \delta_{ij}z,
\quad [d,x_{i}] = x_{i}, \quad [d,y_{i}] = -y_{i} \\ [x_{i},x_{j}] =
[y_{i},y_{j}] = [z,x_{i}] = [z,y_{j}] =[d,z] = 0.\endgather $$ (resp.,
the subalgebra generated by $x_{i}$, $y_{i}$, $z$).

\medpagebreak The quantized universal enveloping algebra
$U_{\hbr}(\hat\g)$ of the extended Heisenberg Lie algebra $\hat\g$ is
the $\C[[\hbr]]$-algebra generated in the $\hbr$-adic sense by
$X_{i}$, $Y_{i}$ ($1\le i \le g$), $Z$, $D$ subject to the relations
$$ [X_{i},Y_{j}] = \delta_{ij} \frac{\sinh \frac{\hbr}{2}
  Z}{\frac{\hbr}{2}}, \quad \text{$Z$ is central,}\quad [D,X_{i}] =
X_{i}, \quad [D,Y_{i}] = -Y_{i}, $$ together with its Hopf algebra
structure, defined by the comultiplication $\De_{\hbr}$, the counit
$\varep_{\hbr}$ and the antipode $S_{\hbr}$: $$\align
\De_{\hbr}(X_{i}) &= X_{i}\otimes \exp(\frac{\hbr}{4}Z) +
\exp(-\frac{\hbr}{4}Z) \otimes X_{i}, \\ \De_{\hbr}(Y_{i}) &=
Y_{i}\otimes \exp(\frac{\hbr}{4}Z) + \exp(-\frac{\hbr}{4}Z) \otimes
Y_{i}, \\ \De_{\hbr}(Z) &= Z\otimes 1 +1 \otimes Z, \\ \De_{\hbr}(D)
&= D\otimes 1 + 1\otimes D,\endalign$$ ${\varep_{\hbr}}_{\vert \langle
  X_{i}, Y_{i}, Z, D \rangle}= 0$, ${S_{\hbr}}_{\vert \langle X_{i},
  Y_{i}, Z, D \rangle}= -\id$. $U_{\hbr}(\hat\g)$ is a
quasi-triangular Hopf algebra and $U_{\hbr}(\g)$ is, by definition,
its Hopf subalgebra generated by $X_{i}, Y_{i}, Z$. The assignment
$$ \aligned \rho(X_i).e_j&=\delta_{j,i+1} e_1,
\\\rho(Z).e_j&=\delta_{j,n+2}e_1, \endaligned \qquad \aligned
\rho(Y_i).e_j&=\delta_{j,n+2} e_{i+1}, \\\rho(D).e_j&= \delta_{j,1}e_1
+ \delta_{j,n+2}e_{n+2},
\endaligned $$ defines a $g+2$-dimensional representation of
$U_{\hbr}(\hat\g)$. Let $\tilde \Gg$ be the connected unipotent
subgroup of $GL(g+2)$ corresponding to $\g$.  The subalgebra of the
dual of $U_{\hbr}(\hat\g)$ generated by the matrix coefficients of
this representation is called the "algebra of rational functions on
the extended quantum Heisenberg group". It can be presented by
generators $\Cal X_i,\Cal Y_i,\Cal D,\Cal D^{-1},\Cal Z$
($i=1,\ldots,g$), with the following relations: $$ [\Cal X_i,\Cal
X_j]=[\Cal X_i,\Cal Y_j]=[\Cal Y_i,\Cal Y_j]=0,\quad [\Cal X_i,\Cal
Z]=\frac{\hbar}{2} \Cal X_i,\quad [\Cal Y_i,\Cal
Z]=\frac{\hbar}{2}\Cal Y_i,\quad \Cal D\Cal D^{-1}=1, $$
($i,j=1,\ldots,n$) and $\Cal D,\Cal D^{-1}$ are central elements
(\cite{Ce et al}, \cite{ALT}). The "algebra of rational functions on
the quantum Heisenberg group" $\Cal A$ is the quotient of the last by
a suitalbe ideal; it is isomorphic to the subalgebra generated by
$\Cal X_i,\Cal Y_i,\Cal Z$. The determination of the Lie algebra
structure on $\g^{*}$ follows easily from \cite{ALT, Prop. 1} (which
generalizes \cite{Dr1, Ex. 2.2}). It turns out that $\Cal A \simeq
U({\g^{*}}_{\hbr})$, where the bracket of ${\g^{*}}_{\hbr}$ is that of
$\g^{*}$ multiplied by $\hbr$.  As ${\g^{*}}_{\hbr}$ is solvable, it
follows from the Kostant-Kirillov "orbit method" (cf. \cite{Di}) that
the primitive spectrum of $\Cal A \simeq U({\g^{*}}_{\hbr})$ is in
natural correspondence with the set of coadjoint orbits of $\g^{*}$.
But we know that the linearization at the identity of the left
dressing action is the coadjoint representation \cite{LW, Th. 2.4}. As
$\g$ is nilpotent, the exponential map is a bijection and therefore
the symplectic leaves of $\hat \Gg$ parametrize naturally the
primitive spectrum of $\A$ (compare with \cite{LS}, \cite{VS},
\cite{HL}, \cite{J}).

 \subheading{\S 3} We are, however, more
interested in quantum versions of another subalgebra of
$U(\g)^{*}$, which will be discussed in this section.

\medpagebreak
If $V$ is a real vector space, we shall denote by
$V[[\hbr]]$ the $\hbr$-adic completion of $V\otimes
\C[[\hbr]]$. It is easy to see that there exists an
isomorphism of $\C[[\hbr]]$-algebras $U_{\hbr}(\hat\g) \to
U(\hat\g)[[\hbr]]$. Therefore we can work with
$U(\g)[[\hbr]]^{*}$.

\medpagebreak
Let $\Gg$ be the real Heisenberg group; as a variety, $\Gg =
S^{1} \times \R^{2g}$; the universal covering of $\Gg$ is of
course $\tilde \Gg(\R)$. Let $(U^{(m)}, \Hc^{(m)})$ be the
Stone--von Neumann representation of the Heisenberg group
$\Gg$ of weight $m$; it is the unique (up to isomorphism)
unitary irreducible representation of $\Gg$ such that
$U^{(m)}_{\la} = \la^{m}\id$ for $\la\in S^{1}$. Moreover, if $(V,
\Hc_{1})$ is any unitary  representation of $G$
such that $V_{\la} = \la^{m}\id$ for  all $\la\in S^{1}$,
then  $\Hc_{1}$ is isomorphic, as an unitary $\Gg$-module,
to $\Hc^{(m)}\hat\otimes \Hc_{0}$, where $\Hc_{0}$ is some
Hilbert space acted upon trivially by $\Gg$. We denote in
particular $\Hc$ instead of $\Hc^{(1)}$.  We shall
apply the preceding to the (diagonal) representation of
$\Gg$ on  $\underbrace{\Hc \hat\otimes \dots \hat\otimes
\Hc}_{m\text{-times}}$, or more generally, to the tensor
product $\Hc^{(m)} \hat\otimes \Hc^{(p)}$, $m, p$ positive
integers.

\medpagebreak
Let $\Hc_{\infty}$ (resp., $\Hc_{-\infty}$) be the
Harish-Chandra submodule of $\Hc$ (resp., its continuous
dual, cf. \cite{Mu, pp. 21 and 24}): $\Hc_{\infty}$ is a
dense subspace of $\Hc$ which carries a representation of
$\g$. Let $v, w \in \Hc_{\infty}$, $\ell, h \in
\Hc_{-\infty}$. Let $\tilde\phi_{\ell, v}: \Gg \to \C$
denote the matrix coefficient $\tilde\phi_{\ell, v}(x) =
\langle \ell, xv \rangle$. We will be mainly concerned with
$\phi_{\ell, v} = \tilde\phi_{\ell, v}s: \R^{2g}\to \C$,
where $s$ is the continuous section $s: \R^{2g}\to \Gg$,
$s(x) = (1,x)$.  In the same vein, consider also the
restriction $\phi_{\ell\otimes h, v\otimes w}$ of a matrix
coefficient of $\Gg$; it is clear that in the algebra of
functions from $\R^{2g}$ to $\C$ the following equality
holds: $$ \phi_{\ell, v}\phi_{h, w} =\phi_{\ell\otimes h,
v\otimes w}. \tag 1
 $$ Alternatively, we can consider  $\tilde\phi_{\ell, v}$ as
an element of $U(\g)^{*}$, and correspondingly, the equality
(1) still holds in this space, which is an algebra via the
comultiplication of $U(\g)$.

It follows from these remarks that the span of the matrix
coefficients $\tilde\phi_{\ell, v}$,  for all $v\in
\Hc^{(m)}_{\infty}$, $\ell \in \Hc^{(m)}_{-\infty}$, $m\in
\Bbb N$, is an associative algebra (without unity). What
happens when replacing the usual comultiplication of
$U(\g)$ by that pushed forward from $U_{\hbr}(\hat\g)$ via
the naive isomorphism alluded above?  The main obstacle is
that the elements of the center of $\g$ are no more
primitive. It follows from \cite{Dr2} that
$U(\hat\g)[[\hbr]]$ is isomorphic (as a Hopf algebra with
the new comultiplication), up to a "gauge" transformation by
an element $F\in U(\hat\g)[[\hbr]] \hat \otimes
U(\hat\g)[[\hbr]]$, to the "standard" quantization of the
pair $(\g, t)$; here $t$ is the invariant symmetric 2-tensor
which arises as the classical limit of $U(\hat\g)[[\hbr]]$.
The explicit isomorphism and $F$ seem to be difficult to
compute; for this reason we introduce in the next sections
new quantized enveloping algebras of $\hat\g$.

 \subheading{\S 4} We define in this section
a new quantization of $U(\g)$ and motivate the
introduction of further quantizations in subsequent
sections.

\medpagebreak Let $\De_I: U(\hat\g)[[\hbr]] \to
U(\hat\g)[[\hbr]]\otimes U(\hat\g)[[\hbr]]$ be the application defined
by $$\align \De_I(x_{i}) &= x_{i}\otimes \exp( \frac{\hbr}{2}z) +
\exp(- \frac{\hbr}{2}z) \otimes x_{i}, \\ \De_I(y_{i}) &= y_{i}\otimes
\exp(- \frac{\hbr}{2}z) + \exp( \frac{\hbr}{2}z) \otimes y_{i}, \\
\De_I(z) &= z\otimes 1 +1 \otimes z, \quad \De_I(d) = d\otimes 1 +
1\otimes d.\endalign$$

\proclaim{Lemma 1} $\De_I$ is well defined and provides
$U(\hat\g)[[\hbr]]$ a Hopf algebra structure, together with the
antipode $S_I$ and the counit $\varep_I$. Its classical limit is the
Lie bialgebra structure on $\hat \g$ given by $$ \delta (x_{i}) =
x_{i} \wedge z, \quad \delta (y_{i}) = -y_{i} \wedge z, \quad \delta
(z) = \delta (d) = 0.\tag 2$$ \endproclaim \demo{Proof}It is
straightforward.\enddemo

\medpagebreak
Recall now a well-known realization of the Stone--von Neumann
representation, and of its Harish-Chandra module
$\Hc_{\infty}$. Let $\upsilon$  be the representation of
$\hat\g$ on the Schwarz algebra on $\R^{g}$ (denoted $\Cal S
(\R^{g})$) given  by  $$\upsilon( x_{i}) f = \pd {t_{i}} f,
\quad \upsilon( y_{i})f = t_{i}f, \quad \upsilon( z)f = f,
\quad\upsilon( d)f = -\sum_{1\le i \le g} t_{i}\pd  {t_{i}}
f. $$
Then  $\Cal S (\R^{g})$ can be identified with
$(\Hc)_{\infty}$.  This $\upsilon$ is the "derivative" of the
representation $U$ of $\Gg$ on $L^{2}(\R^{g})$ given by
$$
U_{\la, y}F(x) = \la\exp(i\pi (2x.y_{2} + y_{1}.y_{2})) F(x+ y_{1}).
$$
$(L^{2}(\R^{g}), U)$ is isomorphic to the Stone--von Neumann
representation of $\Gg$. More generally, the Stone--von Neumann
representation of weight $m$ is $(L^{2}(\R^{g}), U^{(m)})$,
where $U^{(m)}$ is
$$
U^{(m)}_{\la, y}F(x) = \la^{m}\exp(i\pi m(2x.y_{2} +
y_{1}.y_{2})) F(x+ y_{1}). $$
The derivative of $U^{(m)}$ is the representation
$\upsilon^{(m)}$ of $\g$ on $\Cal S (\R^{g})$  given by
$$\gather \upsilon^{(m)}( x_{i}) f = \pd {t_{i}} f, \qquad
\upsilon^{(m)} ( y_{i})f = mt_{i}f, \\ \upsilon^{(m)}( z)f = mf,
\quad\upsilon( d)^{(m)}f = -\sum_{1\le i \le g} t_{i}\pd {t_{i}} f.
\endgather$$ We shall name for brevity $\Hc_{\infty}^{(m)}$ instead of
$(\Cal S (\R^{g}), U^{(m)})$.

Let us now identify $(\Hc)_{\infty}\otimes (\Hc)_{\infty}$
with an  algebra of $C^{\infty}$-functions on $\R^{2g}$.
According with our identification $\R^{2g} \simeq \R^{g}
\times \R^{g}$, we use the variables in $\R^{2g}$ $u_1,
\dots, u_g, v_1, \dots, v_g$.  Then the diagonal action of
$\g$ is given by
 $$\gather \upsilon_0( x_{i})f= (\pd {u_{i}} {}+ \pd {v_{i}}
{})f, \qquad\upsilon_0( y_{i})f = (u_{i} + v_{i})f, \\
\upsilon_0(z)f = 2f, \quad\upsilon_0(d)f = -\sum_{1\le i \le
g} (u_{i}\pd {u_{i}}{} + v_{i}\pd {v_{i}}{})f.
\endgather$$

\medpagebreak

The new coproduct allows us to define "twisted" tensor product of
representations.  In particular, we use $\De_I$ to define, for a fixed
value of $\hbr$, a new representation $\upsilon_I$ of $\hat\g$ on
$\Hc_{\infty} \otimes \Hc_{\infty}$. If $q = \exp( \frac{\hbr}{2})$,
then $$\gather \upsilon_I( x_{i}) f= (q\pd {u_{i}} {}+ q^{-1}\pd
{v_{i}} {})f, \quad \upsilon_I(y_{i})f = (q^{-1}u_{i} +q v_{i})f, \\
\upsilon_I(z)f = 2f, \qquad \upsilon_I(d)f = d.f. \endgather$$

 Notice that there exists a linear isomorphism $\Gc:
\Hc_{\infty} \otimes \Hc_{\infty}\to \Hc_{\infty} \otimes
\Hc_{\infty}$ such that $$ \Gc(\upsilon_0 (x)f) =
\upsilon_I(x)\Gc(f), \tag 3$$ for all $x \in \hat \g, f \in
\Cal S (\R^{g})\otimes \Cal S (\R^{g})$.  Indeed, $$ \Gc (f)
(u, v) = f(q^{-1}u, qv). $$ Note now that $\Gc$ is in fact
well defined as an application $\Gc: L^{2}(\R^{2g}) \to
L^{2}(\R^{2g})$ and hence allows us to define an action $V$
of the Heisenberg group $\Gg$ on $L^{2}(\R^{2g})$ such that
(3) still holds. Explicitly, $$ (V_{\la, y}f)(u, v) =
\la^{2}\exp(2i\pi (q^{-1}u + qv + y_{1})y_{2})f(u + qy_{1},
v + q^{-1}y_{1}). $$ More generally, $\De_I$ allows us to
define new representations $\upsilon_I^{(m,p)}$ of $\hat\g$
on $\Hc_{\infty}^{(m)} \otimes \Hc_{\infty}^{(p)}$:
$$\align \upsilon_I^{(m,p)}( x_{i}) f &= (q^p\pd {u_{i}} {}+ q^{-m}\pd
{v_{i}} {})f, \\ \upsilon_I^{(m,p)}(y_{i})f &= (q^{-p}u_{i} +q^m
v_{i})f, \\ \upsilon_I^{(m,p)}(z)f &= (m+p)f \endalign$$ Again, we
have an intertwining operator $\Gc^{(m,p)}$ between
$\upsilon_I^{(m,p)}$ and $\upsilon_0^{(m)} \otimes \upsilon_0^{(p)}$:
$$ \Gc ^{(m,p)}(f) (u, v) = f(q^{-p}u, q^mv). \tag 4$$ Obviously, $
\Gc ^{(m,p)}$ extends to an application $L^{2}(\R^{2g}) \to
L^{2}(\R^{2g})$ (named identically) and hence we have a representation
$V^{(m,p)}$ of the Heisenberg group $\Gg$ on $L^{2}(\R^{2g})$ by
$$\multline (V^{(m,p)}_{\la, y}f)(u, v) \\ = \la^{m+p}\exp(i\pi
(2mq^{-p}u + 2pq^mv + (m+p) y_{1})y_{2})f(u + q^py_{1}, v +
q^{-m}y_{1}).\endmultline $$ Let now $\ell \in \Hc_{-\infty}^{(m)}$,
$h \in \Hc_{-\infty}^{(p)}$, $f \in \Hc_{\infty}^{(m)}$, $g \in
\Hc_{\infty}^{(p)}$.  Then $\De_I$ provides us a product of
(restriction of) matrix coefficients: $$ \multline\phi_{\ell, f} ._q
\phi_{h, g} (y) = \langle \ell \otimes h, V^{(m,p)}_{1, y}f \otimes
g\rangle = \\ \langle \ell , U^{(m)}_{1,(q^p y_1, q^{-p}y_2)}f \rangle
\langle h, U^{(p)}_{1, (q^{-m} y_1, q^{m}y_2)} g\rangle = \phi_{\ell,
  f}(q^p y_1, q^{-p}y_2) .  \phi_{h, g} (q^{-m} y_1, q^{m}y_2) .
\endmultline\tag 5$$

Notice that the analogue of (5) for $\tilde\phi$ is a new
multiplication on a ring of matrix coefficients of unitary
representations of $\Gg$ of positive weight, that is, an algebra of
functions on $\Gg$.

 \subheading{\S 5} Algebraically, the
multiplication (5) is a special case of the following fact.
Let $\Gamma$ be an abelian group, $A = \oplus_{m\in \Gamma}
A_m$ a $\Gamma$-graded algebra, and $\tau,\sigma:\Gamma \to
\operatorname{Aut} A$ be two representations of $\Gamma$ by
automorphisms of $\Gamma$-graded algebras; suppose in
addition  that $\tau_r\,\sigma_p=\sigma_p\,\tau_r$, for any
$r,p\in \Gamma$.
 We  introduce an associative multiplication $\circ$
on $A$ which  still satisfies $A_{m}\circ A_{p}\subseteq
A_{m+p}$, by the rule  $$
 F\circ G =\tau_p(F)\sigma_m(G),\tag 6
$$ $F\in A_m, G\in A_p$.
It is easy to see that $\circ$ is associative and whit unit
$1\in A_0$. This bi-twisted product is a generalization of
the twisted product defined in \cite{ATV}. There, they
consider $\Gamma=\Bbb Z$, $\tau_p=id$ and
$\sigma_r=\sigma^r$, where $\sigma$ is a $\Bbb Z$-graded
automorphism of a $\Bbb Z$-graded algebra $A$. We denote
this new algebra structure by $A_\sigma$.

\medpagebreak
Here is another example: let again $\Gamma=\Z$, $A$ a $\Bbb
Z$-graded algebra and $\tau$ a graded automorphism of $A$.
Let $\tau_p=\tau^p$ and $\sigma_m=\tau^{-m}$. The resulting
algebra is denoted by $A^\tau$. This is a generalization of
(5).

\proclaim{Proposition 1} If $\sigma=\tau^{-2}$, then
$A_\sigma \cong A^\tau$ via the isomorphism: $\phi: A_\sigma
\to A^\tau$,  defined by $\phi(a)=\tau^m(a)$ for $a \in A_m$.
\endproclaim

The proof is straightforward.

\remark{Remark 1}
Let $X$ be a non-empty set, $M: X \to X$ a bijection, $\k$ a
commutative ring. Let $A = \oplus_{n \in \Z} A_n$, where
$A_n$ is a copy of the algebra of functions from $X$ to
$\k$.  Let $\tau_1(F)(x)=F(M x)$, $\tau_p=\tau_1^p$ and
$\sigma_p=\tau_p^{-1}$.
 If $F\in A_m, G\in A_p$ we have  $$ (F\circ G)(x) =
F(M^px)G(M^{-m}x), \quad x \in X.\tag 7 $$
 This algebra is commutative if and only if $M^{2} = \id$.
This construction extends obviously to any subalgebra of the
algebra of functions on $X$ with values in $\k$, stable by
the transpose of $M$.
\endremark

\subheading{\S 6} The preceding suggests the following construction.
We begin by reversing the reasoning used to obtain (5).  We obtain,
for each pair of positive integers $m$ and $p$, a representation of
$\Gg$ on $L^{2}(\R^{g})$ by the formula $U_{(\la, M^{p}y)}^{(m)
  }\otimes U_{(\la, M^{-m}y)}^{(p)}$ and we seek for an intertwining
operator between it and the usual tensor product representation on
$\Hc^{(m)}\otimes \Hc^{(p)}$. That is, we are looking for an operator
$\Gc$ making commutative the following diagram: $$ \CD
L^{2}(\R^{2g})@>\Gc>> L^{2}(\R^{2g}) \\ @V U_{(\la, y)}^{(m) }\otimes
U_{(\la, y)}^{(p) }VV @VV U_{(\la, M^{p}y)}^{(m) }\otimes U_{(\la,
  M^{-m}y)}^{(p)}V\\ L^{2}(\R^{2g})@>\Gc>> L^{2}(\R^{2g}) .\endCD
$$
Here is a solution: take $ U \in GL(\R^{g})$ and set $M(y) =
(Uy_{1}, \Utr y_{2})$, $\Gc(f)(u,v) = f(U^{-p}u, U^{m}v)$.
Now we conjugate by $\Gc$ the representation of the
Heisenberg Lie algebra on $\Hc^{(m)}_{\infty}\otimes
\Hc^{(p)}_{\infty}$ and  obtain the following formulas as
the derivative of the above representation: $$
\aligned
x_{i} &\quad \text{acts as} \quad \sum_{j}\{(U^{p})_{ji} \pd {u_{j}}
{}  + (U^{-m})_{ji} \pd {v_{j}}{} \}\\
y_{i} &\quad \text{acts as multiplication by} \quad
\sum_{j}\{m(U^{-p})_{ij}  {u_{j}} + p(U^{m})_{ij} {v_{j}} \}\\
z &\quad \text{acts as multiplication by}\quad m+p.
 \endaligned \tag 8$$
 We introduce now a new comultiplication in $U(\g)[[\hbr]]$ which
 explains the preceding representations, but we want first to make
 explicit some straightforward notation. Let $B$ be a matrix in
 $M_{g}(\C)$ and consider the matrix $\hbr z B \in
 M_{g}(\C[[\hbr]][z])$ and consequently the element $\exp(\hbr z
 B)_{ij}$ of $ U(\g)[[\hbr]]$.

\medpagebreak

Let $B \in M_g(\C)$. We define first a Lie bialgebra
$(\g_{B},\delta)$, which is an extension of $\g$ with the additional
property that $\delta(\g)\subset \g\otimes \g$, thus
$(\g,\delta)$ is a sub-bialgebra of   $(\g_{B},\delta)$.

\proclaim{Definition} \rm $(\g_{B},\delta)$ is the Lie bialgebra
generated by $x_i,y_i,z,d$ ($i=1,\ldots,n$), with the following
relations:
$$ [x_i,y_j]=\delta_{ij}z,\quad [d,x_i]= x_i, \quad [d,y_i]= - y_i, $$
$[x_i,x_j]=[y_i,y_j]=0$ and $z$ is central. The structure of Lie
coalgebra is given by $$ \delta (x_{i}) = \sum_{j}B_{ji} x_{j} \wedge
z, \quad \delta (y_{i})= -\sum_{j}B_{ij} y_{j} \wedge z, \quad \delta
(z) = \delta (d) =0.\tag 9 $$ \endproclaim

\remark{Remark 2}  $\g_{B}= \hat\g$ as Lie
algebra.
\endremark

Now we show a quantization of $(\g_{B},\delta)$.

\proclaim{Lemma 2} Let $B \in M_g(\C) $ and let
$\De_{B}:U(\g_{B})[[\hbr]]@>>> U(\g_{B})[[\hbr]] \otimes
U(\g_{B})[[\hbr]] $ be defined by $$ \aligned \De_{B} (x_{i})
&=\sum_{j}\{x_{j}\otimes \exp( \frac{\hbr}{2}zB)_{ji} +
\exp(-\frac{\hbr}{2}zB)_{ji} \otimes x_{j}\}, \\ \De_{B} (y_{i})
&=\sum_{j}\{y_{j}\otimes \exp(- \frac{\hbr}{2}zB)_{ij} +
\exp(\frac{\hbr}{2}zB)_{ij} \otimes y_{j}\} , \\ \De_{B} (z) &=
z\otimes 1+1 \otimes z, \quad \De_{B} (d) = d\otimes 1 + 1\otimes d.
\endaligned \tag 10 $$ It is well defined and provides
$U(\g_{B})[[\hbr]]$ a Hopf algebra structure, together with the
antipode $S$ and the counit $\varep$ defined by $S(u)=-u$ and
$\varepsilon(u)=0$ for all $u \in \g_{B}$. Its classical limit is the
Lie bialgebra structure on $\g_{B}$ given by (9). \endproclaim
\demo{Proof} We denote $U:= \exp(\frac{\hbr}{2}zB)$ and
$\De:=\De_{B}$.  We prove the well-definiteness of $\De $:
  $$
\multline
\De[d,x_i]=\De(x_i)=
 \sum_j \{[d,x_j]\otimes U_{ji} + U^{-1}_{ji} \otimes [d,x_j]\}
= [\De(d),\De(x_i)].
\endmultline
$$ Furthermore
$$ \multline [\De(x_i),\De(y_j)]= \sum_{k,r}\{[x_k,y_r] \otimes
U_{ki}U^{-1}_{jr} + U^{-1}_{ki}U_{jr} \otimes [x_k,y_r]\}\\ = z\otimes
(\sum_k U_{ki}U^{-1}_{jk}) + (\sum_kU^{-1}_{ki}U_{jk}) \otimes z
=z\otimes \delta_{ij} + \delta_{ij} \otimes z =\De[x_i,y_j].
\endmultline
$$ It is possible to verify the other relations in the same way. We
now prove the co-associativity. It relies in the following elementary
remark: if $B_{1}, B_{2}$ commute, $a_1, a_2 \in \Bbb C[[\hbr]]$, then
$ \exp(a_1(z\otimes 1)B_{1} + a_2(1\otimes z)B_{2})_{ij} = \sum_{k}
\exp(a_1zB_{1})_{ik}\otimes \exp(a_2zB_{2})_{kj}$.
$$ \multline (\De \otimes 1)\De (x_i)= \sum_k\{\sum_s x_k\otimes
U_{ks} \otimes U_{si} + \sum_s U^{-1}_{ks} \otimes x_k \otimes U_{si}
+ \sum_s U^{-1}_{ks} \otimes U^{-1}_{si} \otimes x_k\}; \endmultline
$$ and $$ \multline (1 \otimes \De)\De (x_i)= \sum_k\{\sum_s
x_k\otimes U_{ks} \otimes U_{si} + \sum_s U^{-1}_{si} \otimes x_k
\otimes U_{ks} + \sum_s U^{-1}_{si} \otimes U^{-1}_{ks} \otimes x_k\}.
\endmultline
$$ Then we must prove that
 $$
\sum_s
U^{-1}_{ks} \otimes U_{si} =\sum_s U^{-1}_{si} \otimes
U_{ks}\tag 11
$$
 and
$$
 \sum_s U^{-1}_{ks} \otimes
U^{-1}_{si} =\sum_s U^{-1}_{si} \otimes U^{-1}_{ks}. \tag 12
 $$
Now,
$$ \multline \sum_s U^{-1}_{si} \otimes U_{ks}= \sum_s (U^t)^{-1}_{is}
\otimes U^t_{sk} = \exp(\frac{\hbr}2 (-z\otimes 1)B^t
+\frac{\hbr}2(1\otimes z)B^t)_{ik} \\= \exp(\frac{\hbr}2 (-z\otimes
1)B +\frac{\hbr}2(1\otimes z)B)_{ki}= \sum_s U^{-1}_{ks} \otimes
U_{si},
\endmultline
$$ i.e. formula (11) holds.  In analogous way we get (12).

\enddemo

So, we have:

\proclaim{Proposition 2} Let $B \in M_g(\C) $. Consider the Hopf
algebra $(U(\g_{B})[[\hbr]],\Delta_B)$ with a fixed value of $\hbr$
and denote $U:=\exp(\frac{\hbr}{2}B)$. Then the tensor product (via
$\Delta_B$) of $\upsilon^{(m)}$ and $\upsilon^{(p)}$ is given by the
formulas (8) and is denoted by $\upsilon_B^{(m,p)}$.
\endproclaim

\remark{Remark 3} By the same reasoning as above, there exists an
intertwining operator between $\upsilon_B^{(m,p)}$ and
$\upsilon_0^{(m,p)}$. \endremark

 \subheading{\S 7} We defined in \S 6 a family
of new coproducts in $\g$ (parametrized by $B \in M_g(\C)$)
providing  new products in the algebra of matrix
coefficients of unitary representations of positive weight.
In this section, we will see that some of these products
provide a new multiplication on the ring of theta functions.

\medpagebreak
Let $T\in \C^{g\times g}$ be symmetric and such that $\Imm
T$ is positive definite; i.e. $T$ belongs to the Siegel
space $\Hg_{g}$. Let $\cb: \C^{g} \simeq \R^{2g}$ be the
isomorphism provided by $T$. The complex structure on
$\R^{2g}$ provided by $T$ is  $$ J(x_{1}, x_{2})  = \left(
(\Imm T)^{-1}(\re T x_{1} + x_{2}), -\Imm Tx_{1} - \re T
(\Imm T)^{-1}(\re T x_{1} + x_{2})\right)$$ and the
isomorphism is given by $\undx=Tx_{1} + x_{2}
\leftrightarrow (x_{1} , x_{2})$. Let $\Gamma_{m}$ be the
space of holomorphic functions on $\C^{g}$ such that $$ f(z)
= \exp\left(i\pi m (-n_{1}.n_{2} + n_{1}.(2z +\undn)\right)
f(z +\undn). \tag $\theta_{m}$ $$ (Observe that the product
of  functions satisfying ($\theta_{m}$) and ($\theta_{p}$)
respectively satisfies ($\theta_{m+p}$).)

Given a function $f$ satisfying $(\theta_{m})$, we want to
find some $M: \C^{g}\to \C^{g}$ such that  $fM^p$ satisfies
$(\theta_m)$ again ($p \in \Z$). Thus we can apply (7) to
define the bi-twisted product.  An answer is the following.
Let $U\in SO_T(\Z):= O(T) \cap SL(g, \Z)$ (here $O( T)$
means the group of all $U\in GL(\C^{g})$ such that
${}^{t}UTU = T$).  If $T = \Imm T$, the group $SO_T(\Z)$,
being compact and discrete, is finite. Let $M$ be the
translation by $\cb$ of the application $\R^{2g} \to
\R^{2g}$, $x \mapsto (Ux_{1}, \Utr x_{2})$, i.e. $ M:Tx_1
+x_2 \mapsto TUx_1 +\Utr x_2$.

We claim that this $M$ does the job. Indeed, as $U\in SL(g, \Z)$,
$$\multline fM^p(z + \undn ) = f(M^{p}z + M^{p}\undn) = \\
\exp\left(i\pi\{mU^{p}n_{1}. {}^{t}U^{-p}n_{2} - mU^{p}n_{1}. (2M^{p}z
+ M^{p}\undn) \}\right) fM^p(z ).\endmultline$$ We are hence
restricted to show $ mU^{p}n_{1}. (2M^{p}z + M^{p}\undn) \overset ?
\to = m n_{1}.  (2z + \undn) $; but this follows from the requirement
$U\in O(T)$.

\remark{Remark 4} By Proposition 1, we conclude that we
obtain in this way twisted algebras in the sense of
\cite{ATV} (also called Skylanin algebras  \cite{Sk}), via
a quantum multiplication on  matrix coefficients of
Stone--von Neumann representations with positive weight.
\endremark \define\mm{\pmatrix} \define\emm{\endpmatrix}

\remark{Example 1} Let $T=\mm i & 0 \\ 0 & i \emm$, then
$SO_T(\Z)=\left\{ \pm\mm 1 & 0 \\ 0 & 1 \emm, \pm\mm 0 & 1 \\ -1 & 0
\emm\right\}$. So we have two non--isomorphic rings of theta
functions, the classical one and the following: if $f$ satisfies
$(\theta_m)$ and $g$ satisfies $(\theta_p)$, then $f\circ g= (-1)^m
fg$ (this product is obtained via the isomorphism of Proposition 1).
\endremark

 \define\bhq{{\Bbb H}_q} 
\define\bpq{{\Bbb P}_q}  
\define\ott{\underline{\otimes}} 
 \define\ot{\otimes}

\subheading{\S 8} Besides the quantum deformations of the Heisenberg
group considered above, there is another one that has been recently
object of attention: the quantum Heisenberg algebra $\bhq$. This is
the $ \Bbb C[q , q^{-1}] $-algebra generated by elements $X, Y , Z $
with relations
$$
XY - qYX = Z , \qquad  XZ=ZX, \qquad   YZ = ZY.
$$
There is no evident way to define a  Hopf algebra structure
on $\bhq$. We shall show that however the situation is
different if one considers  a natural twisted algebra
structure in the tensor product.

Let $(A,m,1)$ be an algebra. Let $S: A\otimes A \to A\otimes A$ be an
invertible linear transformation such that \roster
\item"({\it i})" $S^{12}S^{23}S^{12}=S^{23}S^{12}S^{23}$, i.e. $S$ is
  a solution of the braided relation (sometimes called the quantum
  Yang--Baxter equation).

  \item"({\it ii})" $S(b\otimes 1)=1\otimes b$ and $S(1\otimes
    b)=b\otimes 1$.

    \item"({\it iii})" $S(m\otimes id)=(id \otimes m)S^{12}S^{23}$ and
      $S(id \otimes m)=(m \otimes id)S^{23}S^{12}$.

      \item"({\it iv})" $S^2=1$ \endroster The pair $(A,S)$ is called
        a {\it braided algebra}. Suppose that in addition \roster
        \item"({\it v})" $mS=m$. \endroster Then the pair $(A,S)$ is
          called a {\it $S$-commutative} or {\it generalized
            commutative} or {\it braided commutative algebra}. We
          prefer this last term. See \cite{GRR}, \cite{Ma}, \cite{Mn}.

\bigpagebreak
The following example was found by Demidov (see \cite{Mn})
and is known as the ``quantum plane" (notice that related
skew-fields are known to algebraists since long time ago).
Let $ \bpq$ be the quotient of the tensor algebra $T(V)$
(where $V$ is a 2-dimensional vector space with basis $X,Y$)
by the 2-sided ideal generated by $ X\ot Y - q Y\ot X $. Now
let $I$ be the ideal of $\bpq \ot \bpq$  generated by $
X\ot Y - q Y\ot X$.  Let $S: \bpq\otimes \bpq \to \bpq\otimes
\bpq$ be  the unique linear application  such that $$
X\otimes X \mapsto X \otimes X ,\quad X\otimes Y \mapsto qY
\otimes X,\quad Y \otimes X \mapsto q^{-1}X \otimes Y,\quad
Y \otimes Y \mapsto Y \otimes Y,
$$
 and $S$ preserves $I$. $S$ is a linear isomorphism;
moreover $(\bpq,S)$ is easily seen to be braided commutative.
Observe that $
S(X^nY^p \otimes X^m Y^r) = q^{nr-pm} (X^mY^r \otimes
X^nY^p)$. (It is known that  $\bpq$ generalizes  to
higher dimensional $q$-affine  spaces; the definition of $S$
and the following results  are also valid for them).

The following Lemma was communicated to the first author by P. Cartier
(see also \cite{GRR}, \cite{Ma}, \cite{Mn}).  \proclaim{Lemma 3} If
$(A,S)$ is a braided algebra then $A\otimes A$, with the product: $$
(a\otimes b)*(c\otimes d)= (m \otimes m)(a \otimes S(b \otimes
c)\otimes d)
$$ is an associative algebra with unit $1\otimes 1$. It will be
denoted by $A \ott A$.  \endproclaim \demo{Proof} Let $a,b,c,d,e,f \in
A$ and denote $d_i \ot e^i=S(d\ot e)$; then
$$\multline
(a \ot b)*((c\ot d)*(e\ot f))= (m\ot m)(a \ot S(b\ot
cd_i)\ot e^if) = \\ (m \ot m)(id \ot m \ot id \ot id)(a \ot
S^{23}S^{12} (b \ot c \ot d_i) \ot e^if)= \\ (m \ot m)(id
\ot m \ot id \ot m)(a \ot S^{23}S^{12} S^{34} (b \ot c \ot d
\ot e) \ot f).
\endmultline
$$ In analogous way we can show that:
$$
((a \ot b)*(c\ot d))*(e\ot f)=(m \ot m)(m \ot id \ot m \ot
id) (a \ot S^{23}S^{34} S^{12} (b \ot c \ot d \ot e) \ot f).
$$
Using the associativity of $m$ is enough to prove that:
$$
S^{23}S^{12} S^{34}=S^{23}S^{34} S^{12},
$$
and this is true because $S^{12} S^{34}=S^{34} S^{12}$.
\enddemo

Notice that $\bpq \ott \bpq$ is isomorphic to the 4-dimensional
$q$-affine space.

\proclaim{Lemma 4} There exists a unique algebra
homomorphism $\Delta : \bpq @>>> \bpq \ott \bpq $ such that
$$ \Delta (X) = X \ott 1 + 1 \ott X, \qquad \Delta(Y) = Y \ott 1 + 1
\ott Y.\tag 13 $$ Moreover $(\Delta \ott id)\Delta = (id \ott
\Delta)\Delta$. That is, $\bpq$ is a braided bialgebra (the counit
$\varepsilon$ is defined by $\varepsilon(X)=\varepsilon(Y)=0$).
\endproclaim

Now we pass to $\bhq$. Let $\tilde S: \bhq \otimes \bhq
\to \bhq \otimes \bhq$ be defined by
$$
\tilde S: X^nY^p Z^t \otimes X^m Y^r Z^v \mapsto
q^{nr-mp} X^mY^r Z^v \otimes X^n Y^p Z^t. \tag 14
$$

\proclaim{Proposition 3} ({\it i})$(\bhq, \tilde S)$ is a
braided algebra and $\bhq \to \bpq$, $Z \mapsto 0$ is a
morphism of braided algebras.

({\it ii}) Let $\Delta : \bhq @>>> \bhq \ott \bhq $ be defined by (13)
and $\Delta(Z)=1\ott Z + Z \ott 1$. Then $(\bhq,\Delta)$ is a braided
Hopf algebra (the counit takes the $0$ value in $X,Y,Z$).  Let
$\bhq^{op}$ be $\bhq$ with the multiplication $m\tilde S$, when $m$ is
the multiplication. Then there exists a unique morphism of algebras
$\Cal S: \bhq \to \bhq^{op}$ such that $\Cal S(X)=-X$, $\Cal S(Y)=-Y$
and $\Cal S(Z)=-Z$. Is easy to verify that $\Cal S$ is the antipode of
$\bhq$.  \endproclaim

One has thus an
exact sequence of braided Hopf algebras
 $$ 0 @>>> k[Z] @>>> \bhq @>>> \bpq @>>> 0.  $$

\heading Acknowledgments
\endheading

The work of the first author was supported by the Alexander von
Humboldt-Stiftung and was done at the Max Planck Institut (Bonn). We
also want to thank P. Slodowy for his interest on this work; the third
author acknowledges the invitation of P. Slodowy to Bonn in November
92, when the seminal ideas of this paper were found at.

 \enddocument